\documentclass[twocolumn,amsmath,amssymb]{revtex4}
\usepackage{amsmath}
\usepackage{graphicx}
\usepackage{color}
\usepackage{mathtools}
\usepackage{multirow}
\usepackage{bigints}

\begin{document}

\preprint{APS/123-QED}

\title{Self-similarity with universal property for soap film and bubble in roll-off regime}

\author{Wei-Chih Li, Chih-Yao Shih, Tzu-Liang Chang and Tzay-Ming Hong$^{\dagger}$}

\affiliation{Department of Physics, National Tsing Hua University, Hsinchu 30013, Taiwan, Republic of China}

\collaboration{CLEO Collaboration}

\date{\today}

\begin{abstract}
All children enjoy blowing soap bubbles that also show up in our bath and when we wash dishes. We analyze the thinning and breaking of soap bubble neck when it is stretched. To contrast with the more widely studied film whose boundaries are open, we concentrate on the bubble with a conserved air volume $V$. Like film (F), non-equilibrium state can  be divided into four regimes for bubble (B): (1) roll-off, (2) cusp approach, (3) pinch-off and (4) breakup. We establish the existence of  self-similarity in  F-1, B-1 and B-3, and universal property in  F-1 and B-1 for the profile of soap membrane. The former means that the  profile at successive times can be mapped to a master curve after being rescaled by the countdown time $\tau$. Whiles, the latter further requires this master curve to be identical for different ring size $R$ for film and different $V$ and $R$  for bubble while keeping $V/R^3$ fixed.  The exhibition of universal property indicates that the process of memory erasing starts earlier than regime 3. We also found that the  minimum radius scales as $h_{\rm min}\sim \tau ^{1/2}$, independent of $V$ and pulling speed.    Note that the validity of our discussion is limited by the duration of roll-off regime from $10^{-2}\sim 10^{-3}$  s. 

\end{abstract}

\maketitle

\section{Introduction} 
\begin{figure*}
	\centering	
	\includegraphics[width=18cm]{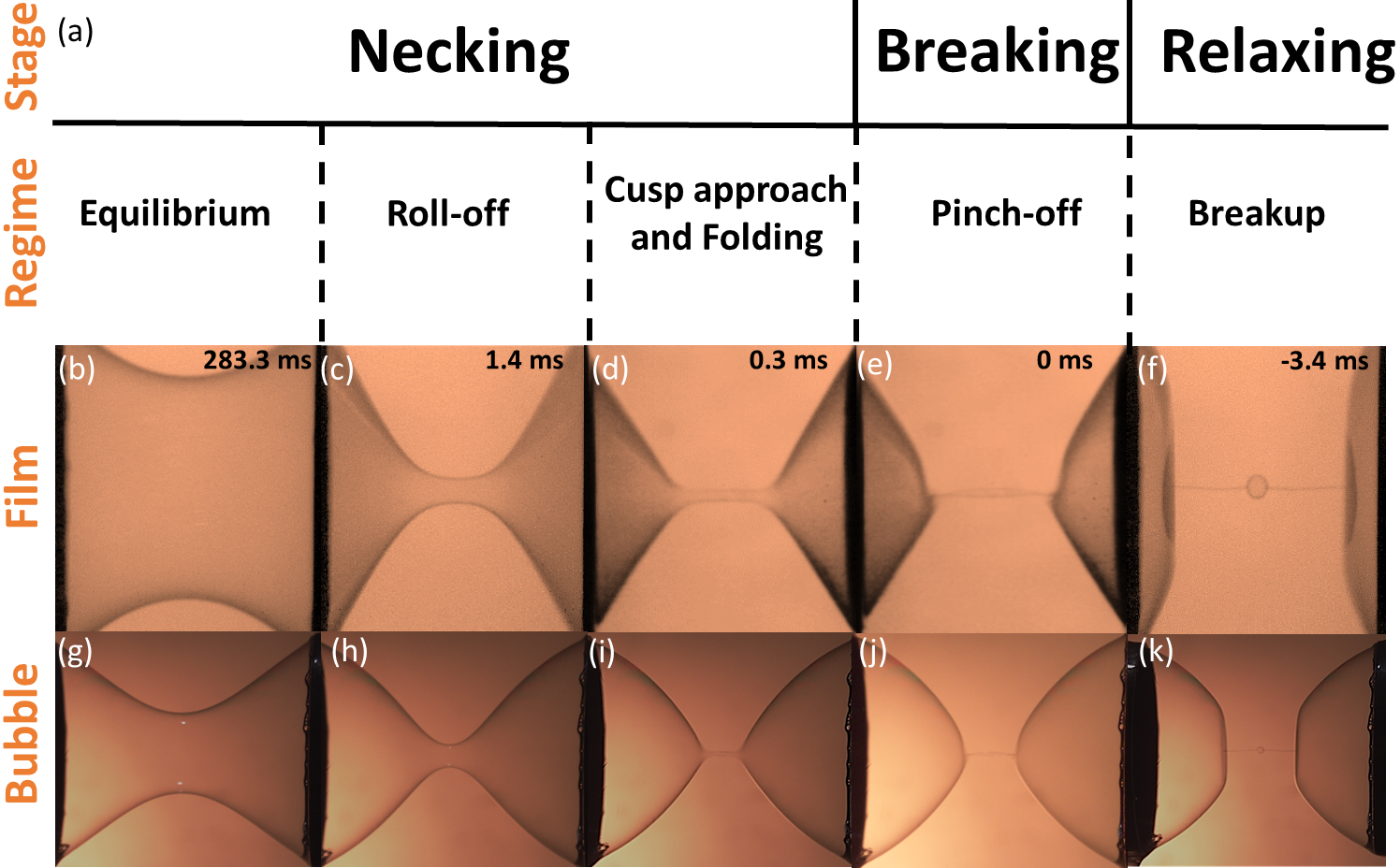}
	
    \caption{(a) The procedures of breakage consist of three stages  which can be further divided into five regimes. Their corresponding photos at different $\tau$ for
  the soap film and bubble are shown in (b$\sim$f) and (g$\sim$k), separately.  Note that the film is allowed to squeeze air out of its interior, while the bubble has to roughly conserve its volume. The pulling speed  is set at $v_{s}=$ 16 mm/s, the radius of ring or cap $R=$ 20 mm, and the pumping volume of bubble $V=$ 26 ml. }
    \label{fig:bubble_set}		    
\end{figure*}

In addition to academic interests on the dynamics leading to the formation of singularities  and its practical importance in industrial processes such as ink-jet printer\cite{material_printing} and injection moulding\cite{molding},  the breakup of  a fluid body into two or more pieces may even shed  light on the understanding of  cell division \cite{cell} of animals. 
Its procedures have been roughly divided into three stages\cite{Chen_steen_numerical} - necking, breaking and relaxing according to Steen {\it et al.}. They are further subdivided into six regimes: equilibrium, roll-off, cusp approach, folding, pinch-off and breakup. To help clarify the nomenclature, we line up these phrases with their corresponding photos in Fig. \ref{fig:bubble_set}  for soap membranes with open or close boundaries which we shall call the film or bubble from now on.  The first four regimes belong to the necking stage, while  pinch-off and breakup regimes refer separately to the breaking and relaxing stages. We shall merge  cusp approach and folding  into one regime because, like the use of superfluid by Burton {\it et al.} \cite{superfluid}, we were not able to distinguish them in our nearly inviscid soap experiments. 

The phenomenon of pinch-off occurs in the breaking stage and is mostly dominated by three major mechanisms\cite{Hydrodynamic_stability,Breaking_of_liquid_films_and_threads,slender,bir_bridge,
Universal_Eggers,bir_pendant_drop,
Iterated_Instabilities_nagel,Drop_formation_eaggers,inital_nagel,
droplets_from_liquid_jet,Drop_formation_in_viscous_flows,Lister_H_stone,Computational_experimental_drop_formation,Satellite_drops_pinch_off,
Drop_formation_from_a_capillary_tube_one_d,Modeling_pinchoff_and_reconnection,Ligament_Mediated_Spray_Formation,
Scaling_and_Instabilities_Bubble,
giant_bubble_pinch,
bubble_nagel,
Simplicity_and_complexity_in_dripping,Computational_analysis_of_drop,pinch_rewiew,
Pinching_Dynamics_and_Satellite_Droplet_Formation_in_Symmetrical_Droplet_Collisions,yc, two_fluid_snap,self_differ_viscosity,
Testing_for_scaling,self_viscosity_thm}: surface tension, inertia and viscosity. The minimum radius $h_{\rm{min}}$ has been found to decrease as $\tau^{2/3}$ where  $\tau \equiv t_{p}-t$,  the countdown time, is defined as the difference between  the pinch-off time  $t_{p}$ and the real time $t$, when viscosity is negligible in systems such as water dripping in the air. Similar to the dominant role of surface tension and inertia, both water and air can be approximated as being nearly inviscid. In the meantime, the pinch-off of inviscid fluid system  has been studied by theoretical, experimental and computational analyses \cite{Self_Similar_Capillary, C_and_E_scaling, Capillary_pinch_off, mercury,EPL_film,superfluid,self_inviscid_exp}. Although being the driving force for all  the above cases \cite{Self_Similar_Capillary, C_and_E_scaling, Capillary_pinch_off, mercury, superfluid, two_fluid_snap,Testing_for_scaling},  the  surface tension is not necessary to render the pinch-off, e.g.,  thermal fluctuation or bulk diffusion \cite{thermal_fluctuation, diffusion} is known to be equally capable as the capillary force is ultra-low.

The final breakage of liquid drop into several pieces is universal in most cases, meaning that its collapsing speed and neck radius are independent of initial and boundary conditions\cite{Universal_Eggers, Iterated_Instabilities_nagel, Drop_formation_eaggers, droplets_from_liquid_jet, Ligament_Mediated_Spray_Formation, Pinching_Dynamics_and_Satellite_Droplet_Formation_in_Symmetrical_Droplet_Collisions,two_fluid_snap,self_differ_viscosity,Testing_for_scaling}. In other words, the memory is erased. Targeting the particular case of a water droplet in silicon oil, Nagel {\it et al.} \cite{memory} found that both the sine wave perturbation in numerical simulation and  different boundary conditions in experiments would affect the curvature at the pinch-off neck. A similar system, for which the viscosity of the inner fluid is much smaller than the outer one, was investigated by Stone {\it et al.}\cite{universal_stone} and found to lose its memory again when the outside container is confined to  a capillary tube.     

Open soap film belongs to inviscid fluid and forms a catenoid  to minimize its surface tension energy\cite{minimal_surface_Euler_Leonhard,Plateau, A_Complete_Minimal_Surface_in_R3_Between_two_Parallel_Planes,Shapes_of_embedded_minimal_surfaces}. For instance, the numerical work \cite{Chen_steen_numerical} by Chen and Steen  fits the $h_{\rm{min}}(\tau )$ by a power law with an exponent $2/3$ in  pinch-off regime. Subsequently, Robinson and Steen \cite{robinson_steen_exp,cryer_steen_exp}  did some experiments to verify  their previous numerical result  \cite{Chen_steen_numerical}. Limited by the frame rate of their high-speed camera, they were not able to verify the property of self-similarity, that has been observed in bubble bursting\cite{bubble_yao}, relaxation of confined droplets\cite{relaxation} and liquid lens coalescence\cite{lens}, for pinch-off regime. 

Compared with soap film, it is apparent that bubble looks plumper in Fig. \ref{fig:bubble_set}. One reasonable question  is whether the  constraint of volume conservation will sustain to  alter the fascinating dynamics of $h_{\rm{min}}$ and profile when the system enters non-equilibrium state.   
 In equilibrium regime, the shrinking of the neck is quasi-static and reversible, i.e.,  the neck radius can retain its original value when the separation distance $L$ reverts to the previous length. When extended to a critical length $L^{*}$,  both film and bubble become unstable and their time evolution is shown in Fig. \ref{fig:bubble_set} and characterized by: (1) roll-off with only one $h_{\rm {min}}$, (2) pinch-off when two necks suddenly emerge symmetrically with a separation distance that increases much slower than the collapsing speed of necking and (3) breakup at which film and bubble separate after pinching-off. Besides clarifying the role of long-range medium pressure  due to volume conservation for bubble, we also want to investigate whether the self-similarity is exhibited not only in the final pinch-off for bubble; but also in roll-off regime.
Not limited to bubble, the volume of ink droplet is also conserved and how it affects  $L^*$ has incurred some debates  in the printing industry\cite{Simulation_ink_gravure}. Previous researchers \cite{Stability_of_liquid_bridges_between_equal_disks_in_an_axial_gravity_field,vertical_abtrary_liquid_bridge, 
On_the_breakup_of_viscous_liquid_threads,nonlinear_bridge} studied the breakage of liquid that was stored between the gap of two departing horizontal and vertical rods. Although they found the volume of liquid to affect $L^*$, their conclusions were tainted by the movement of the contact line\cite{contact_line} when the neck collapses. In contrast, our setup, as described in the next section, is free of such a defect while honoring the volume-conservation.

This paper is organized as follows: Experimental setup  is described and relevant parameters are defined in Sec. II. How the neck radius varies with the separation distance is studied for  equilibrium regime and compared between film and bubble in Sec. III A.  While the self-similarity and  universal property are generally thought to be unique for  pinch-off regime in Sec. III C, we check and confirm their existence in as early as roll-off regime
 in Sec. III B, where the power-law relation $h_{\rm{min}} \sim \tau^{\alpha}$ is also examined. The  final breakup regime that is characterized by the formation of a satellite bubble is arranged in  Sec. III D, where data on how the pumping volume $V$ affects $L^*$ are presented.  Complementing the experimental results in Sec. III, theoretical models and derivations are presented in Sec. IV.   Finally, we conclude and suggest possible directions for future workers in Sec. V.

\begin{figure}
	\centering
	\includegraphics[width=8.5cm]{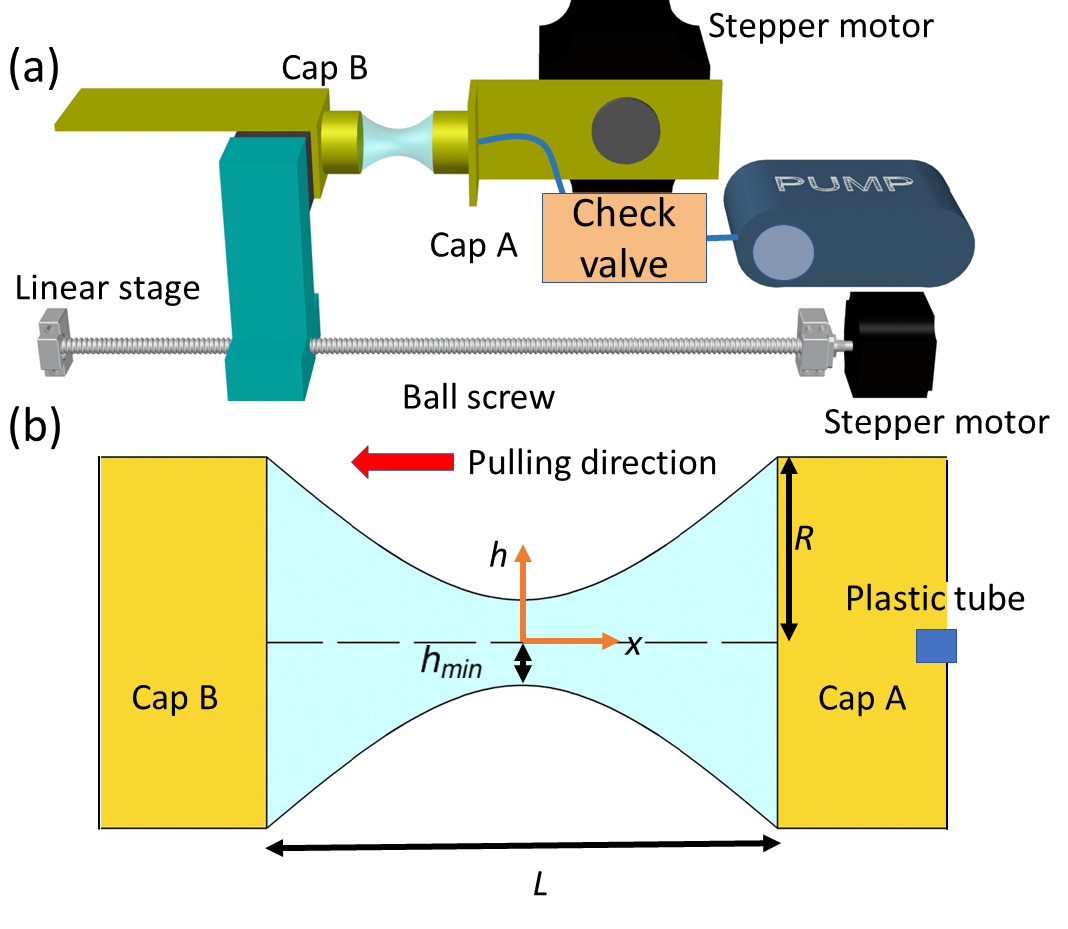}

    \caption{(a) Schematic experimental setup for stretching soap bubble by a stepper motor. (b) Relevant parameters are defined. Bubble is painted in blue, while the caps are  in yellow. Cap A and B are replaced by Ring A and B to produce film.}
    \label{fig:set_flat}
\end{figure}

\section{Experimental setup}
The ingredients \cite{Chen_steen_numerical} of our soap water include dried oleic acid soap, deionized water and guar gum. The addition of guar gum has been verified\cite{big_bubble} to prolong the lifetime of soap membrane. After dipping into the solution, an aluminum cap A of radius $R$ is rotated 90 degrees by a stepper motor to horizontally align its open end to another cap B at a distance $L$. When the air pump is switched on by a solid state relay module, a soap bubble is formed on cap A. We use a check valve to ensure that no back-flow of air will deflate bubble. As demonstrated schematically in Fig. \ref{fig:set_flat}, this bubble is gently attached to cap B that is pre-wetted. Then, a linear ball screw driven by another stepper motor is utilized to move cap B away from cap A with a constant pulling speed $v_{s} \approx 16  $ mm/s. Collapse of bubble neck is recorded by a high speed camera with 23000 fps. We originally open a hole on cap B to change the bubble to film. However, the film shape is always plagued by an asymmetry due to the time lag it takes for the air to flow out. As a result, the caps are eventually replaced by two rings to generate film. 

The dominated term in the collapse is determined by several dimensionless numbers. To begin with, we deduce that the shear viscosity is negligible since Reynolds number Re= $\rho \bar{v_{c}} \bar{h} / \eta \approx 10^{2}\gg 1$ where notations are defined in Table \ref{parameter}.  In the mean time, the magnitudes of the other two numbers, Bond number Bo=$\rho g R / \gamma \approx 10^{-2}$ and Webber number We = $\rho \bar{v_{c}}^{2} \delta / \gamma\approx 10^{-1}$, reassure us  of two things. First, soap bubble can be regarded as being symmetric since the effect of gravity is small. Second, the system is dominated by the surface tension and inertia. Surface tension  coefficient is estimated by $\gamma =\rho g( 6V_{d}/\pi )^{2/3}$\cite{surface_tension_drop}. 
\begin{table}[ht]
\caption{ Notation and definition for relevant parameters}
\begin{tabular}{|c|c|}
\hline
\multicolumn{1}{|c}{Notation} & \multicolumn{1}{|c|}{Definition}               \\ \hline
$L$                            & separation distance between rings or caps                    \\ \hline
$R$                            & radius of  ring or cap                     \\ \hline
$t$                            & real time                           \\ \hline
$L^{*}$                          &critical length \\ \hline
$V$                            & pumping volume of air                           \\ \hline
$h$                            & radius of cross section                           \\ \hline
$\bar{h}$                            & characteristic radius of cross section                           \\ \hline
$h_{\rm{min}}$                            & minimal radius at neck \\ \hline
$\delta$        & thickness of film or bubble                                \\ \hline
$\gamma$                            & surface tension coefficient                    \\ \hline
$\bar{v_{c}}$                            & characteristic collapse speed                   \\ \hline
$\rho$                            & mass density of soap water                   \\ \hline
$v_s$                            & pulling speed                   \\ \hline
$V_d$                            & volume of soap droplet                                 \\ \hline
$\sigma$        & surface mass density of  film or bubble          \\ \hline
$\tau$          & difference between pinch-off and real times                                      \\ \hline
$t_p$       &  pinch-off time             \\ \hline  
$x_p$       &  pinch-off position in $x$ axis            \\ \hline  
$\eta$       &  shear viscosity of soap water             \\ \hline  
$\lambda$       &  Lagrange multiplier             \\ \hline    
$P$       &  pressure difference across bubble surface        \\ \hline 
$\kappa_1, \kappa_2$       &  radial and axial curvatures        \\ \hline

\end{tabular}
\label{parameter}
\end{table} 
\begin{figure}[h!]
    \centering
    
	\includegraphics[width=8.7cm]{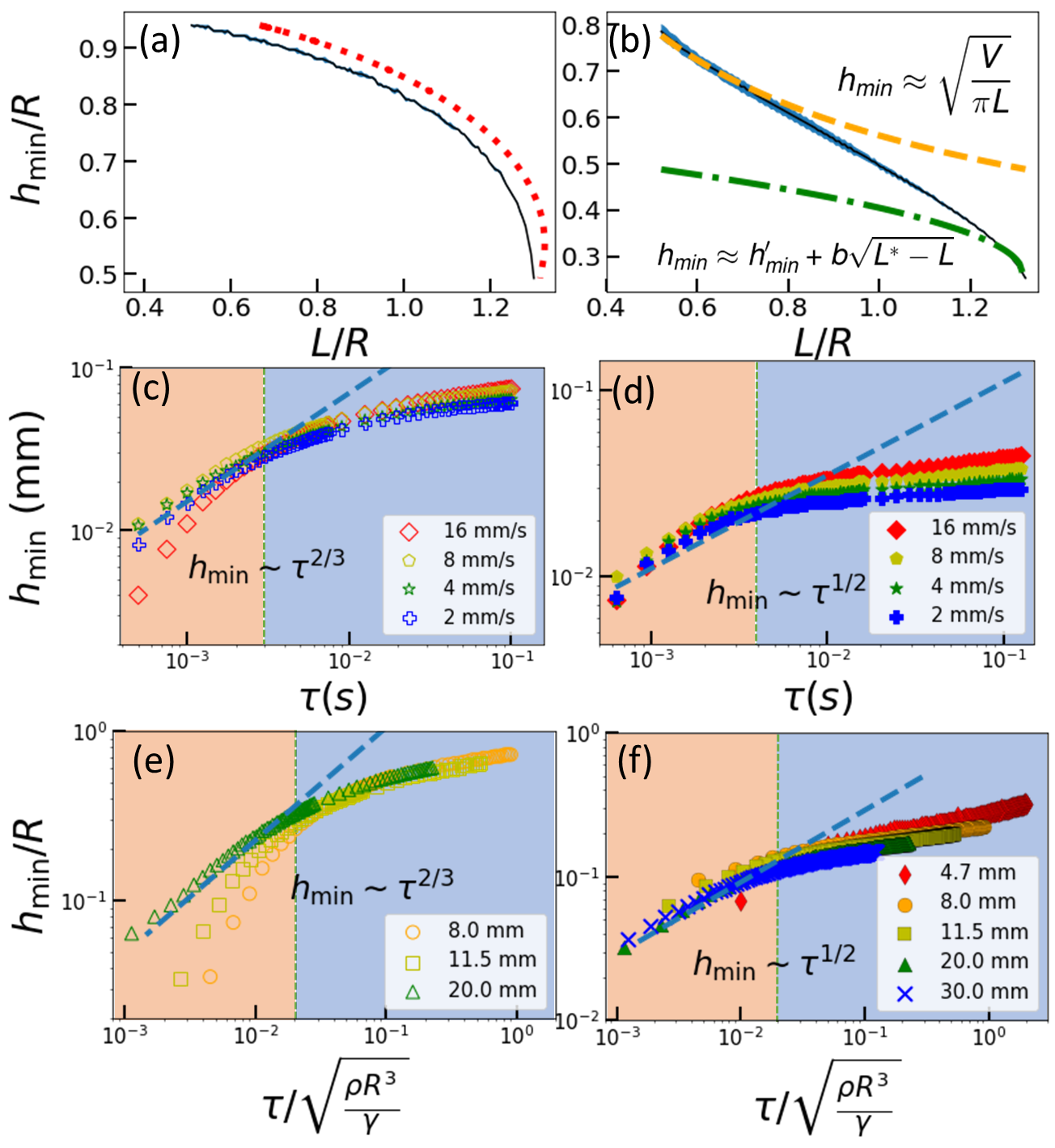}   
    \caption{ Normalized $h_{\rm{min}}/R$ vs. $L/R$ for (a)  film and  (b) bubble whose neck shrinks during equilibrium regime where $R=$  20 mm, $v_{s}=$  16 mm/s and $V=$ 26 ml. Unnormalized $h_{\rm{min}}$ as a function of $\tau$ at different $v_{s}$ for (c) film and (d) bubble where $R=$ 11.5 mm and $V=$ 2.9 ml.  (c, d) are rescaled in (e, f) at different $R$ where $v_{s}=$ 16 mm/s and $V/R^3=$ 3.2. Theoretical predictions are in dotted, dashed and dash-dotted lines.}
    \label{fig:sp_volume}
\end{figure}

\section{Experimental results}
\subsection{Equilibrium regime}
Compared to film, the volume of bubble roughly remains constant  when pulled apart, as argued in Sec. IV of the Supplemental Material (SM) \cite{sm}. The center of  bubble surface evolves from being convex to concave in contrast to  film that is always convex throughout  equilibrium regime. This difference will be argued later in the theory section to give rise to a positive second derivative of $h_{\rm{min}}(L/R)$ in bubble, as opposed to a negative one in  film \cite{Chen_steen_numerical, robinson_steen_exp,cryer_steen_exp}, as shown in Fig.~\ref{fig:sp_volume}(a, b). When $\gamma$, $R$ and $V$ are fixed, $h_{\rm{min}}$ can be uniquely determined by $L$ in equilibrium regime. By use of $L^*-L = v_{s} (t^*-t)$ where $t^*$ denotes the time when the neck starts to collapse spontaneously, $h_{\rm{min}}$ can be alternatively expressed as a function of $v_{s}(t^*-t)$. But, since  non-equilibrium state proceeds much faster than $v_{s}$, the time it takes is negligibly small, i.e., $t_p - t^*\approx 0$. So, we can use $\tau \equiv t_p-t$ to track the evolution of $h_{\rm min}$ from now on. 

The necking process is expected to be size-independent for both film and bubble. Therefore we divide $h_{\rm{min}}$ and  $\tau$  by $R$ and $R/v_{s}$ to render them dimensionless. When $\tau$ is fixed, $L$ is proportional to $R$ at the same $v_{s}$  for a film. A different choice of ratio of $V/R^{3}$ will shift the curve in Fig. ~\ref{fig:sp_volume}(f) that still remains independent of $R$, as detailed in Sec. VII of the SM \cite{sm}. 
   
\subsection{Roll-off regime} 
Upon entering  roll-off regime, film and bubble share the following properties: First, the evolution of $h_{\rm{min}}$ is independent of (1)  the pulling speed $v_{s}$ in Fig.~\ref{fig:sp_volume} (c, d) because the shrinkage is much faster and (2) the ring or cap size if  both  $h_{\rm {min}}$  and $\tau$ in Fig.~\ref{fig:sp_volume} (e, f) are properly rescaled by $R$  and  the characteristic time $ \sqrt{\rho R^{3}/\gamma}$ from  the balance between the surface tension and inertial force. Note that the $h_{\rm min} (\tau )$ for film has been predicted to be independent of $R$ by \cite{Chen_steen_numerical}. What distinguishes bubble from film is that  $\alpha =1/2$ for  the former instead of 2/3 \cite{Chen_steen_numerical}.   Although $h_{\rm{min}}$ will increase with more pumping volume,  the value of  $\alpha$ is checked to be independent of $V$, as shown in Fig \ref{fig:volume_appendix_tot}. We believe this is due to the fact that there is little air in the vicinity of bubble neck. However, the constraint imposed by the volume conservation is still critical at modifying $\alpha$. Heuristically the existence of a pressure difference across the membrane impedes the collapsing of bubble and thus renders a small $\alpha$.

\begin{figure}[h]
	\includegraphics[ width=9cm]{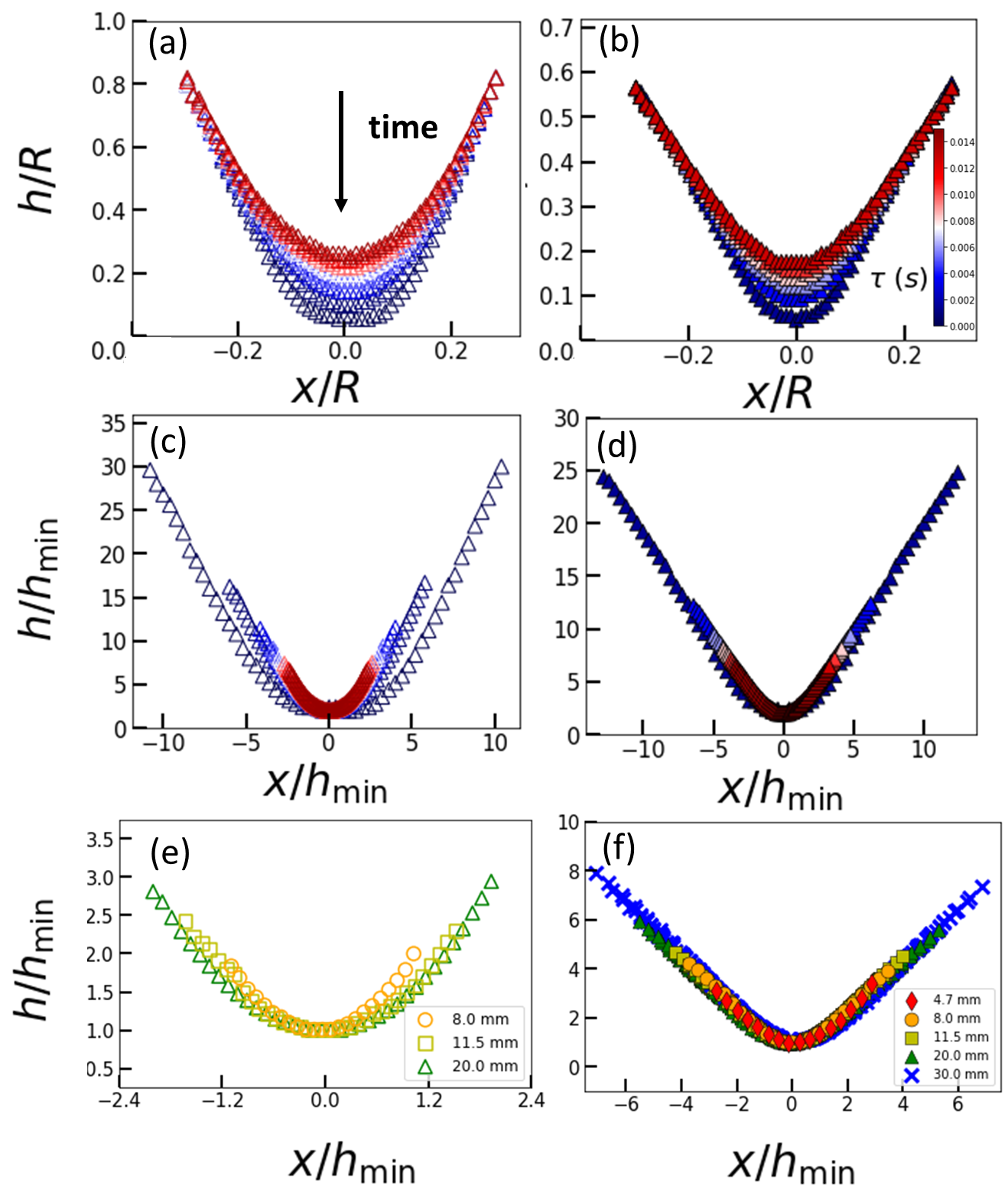}
    \caption{Evolution for the profile of (a) film and (b) bubble in roll-off regime where $\tau$ is denoted by different colors. When  rescaled by $h_{\rm{min}}$ in  (c, d), their contours are found to  follow self-similar behavior, whose
 master curves are checked to be independent of the ring or cap size  in (e, f).}
    \label{fig:oc_power1}        
\end{figure}

The second property shared by film and bubble regards the evolution of profile, $h(x)$ in Fig.~\ref{fig:oc_power1} (a, b) and includes: (1) Both are found to exhibit self-similarity, i.e., their shapes at different time can be mapped to a master curve if both $h$ and $x$ are rescaled by $h_{\rm{min}}$, as shown  in Fig.~\ref{fig:oc_power1} (c, d). The level of similarity is quantified by  cosine similarity \cite{self}  = 0.98 and 0.99  for  film and bubble, as detailed in Sec. II of the SM \cite{sm}.  (2) Both master curves do not change with  different ring or cap size in  Fig.~\ref{fig:oc_power1}(e, f).  In contrast to  $h_{\rm{min}}$ that is independent of both $R$ and $V$, the contour in Fig.~\ref{fig:oc_power1}(f) is only universal with respect to  $V/R^{3}$. When  $V/R^{3}$ changes, the contour will become different. What distinguishes bubble from film is that $h''(x=L/2)<0$ which renders an inflection point between $x=0$ and $x=L/2$, where $h''$ denotes $d^2h/dx^2$. In contrast, the film is always concave upward. This is demonstrated theoretically in Sec. X of the SM \cite{sm}.

\subsection{Pinch-off regime}
\begin{figure}[h]
	\includegraphics[ width=9cm]{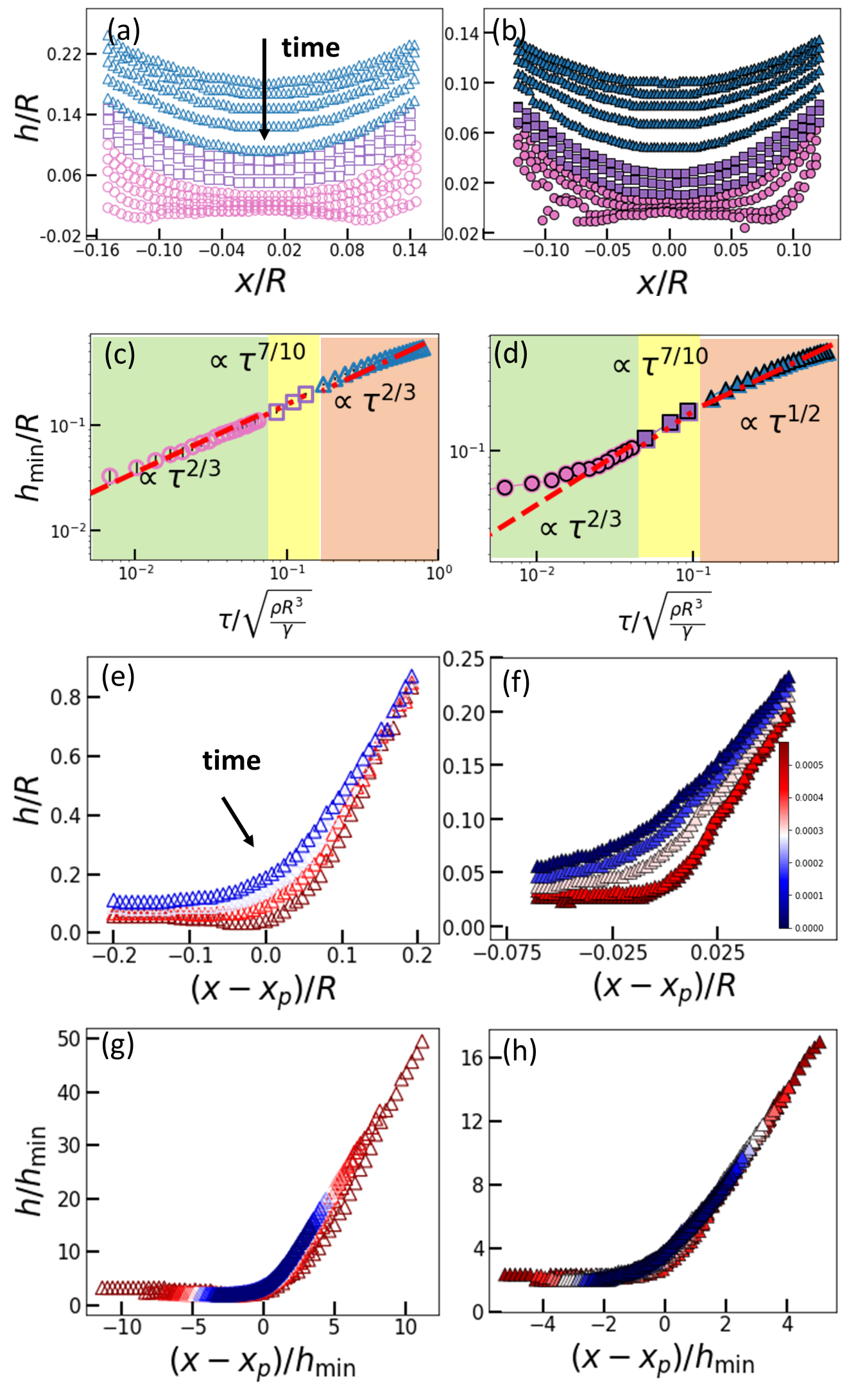}

    \caption{Figure \ref{fig:oc_power1}(a, b)  are extended to include  pinch-off regime in (a, b). Similar expansion is done for Fig. \ref{fig:sp_volume}(e, f) to obtain (c, d) where roll-off, cusp approach and pinch-off regime are separately denoted by triangles, squares and circles in (a$\sim$d). Time evolution of  (e) film and (f) bubble in pinch-off where $\tau$ is denoted by different colors and $R=$ 20.0 mm, $v_{s} =$ 16 mm/s and $V$= 26 ml. Only when (e, f) are rescaled by $h_{\rm min}$, do the shapes reveal the property of self-similarity in (g) and (h).}  
    \label{fig:oc_power2}        
\end{figure}

The number of neck can be seen to double as film and bubble transits from roll-off to pinch-off regime in Fig.~\ref{fig:oc_power2} (a, b). Furthermore, the originally different exponent $\alpha$ in roll-off becomes identical at 2/3 for film and bubble  near pinch-off regime in   Fig.~\ref{fig:oc_power2} (c, d),  rescaled  as  Fig. ~\ref{fig:sp_volume} (e, f). 
The evolution of profile for film looks similar to that of bubble in pinch-off regime with the shifting of minimum point from $x=$0 to $x=x_{\rm p}$ in Fig. \ref{fig:oc_power2} (e, f).  The self-similar behavior which has been predicted by numerical and theoretical works \cite{Chen_steen_numerical, Self_Similar_Capillary} for film turns out to be also true for bubble, as shown in Fig. \ref{fig:oc_power2} (g, h).

\begin{figure}[h!]
	\includegraphics[ scale=0.42]{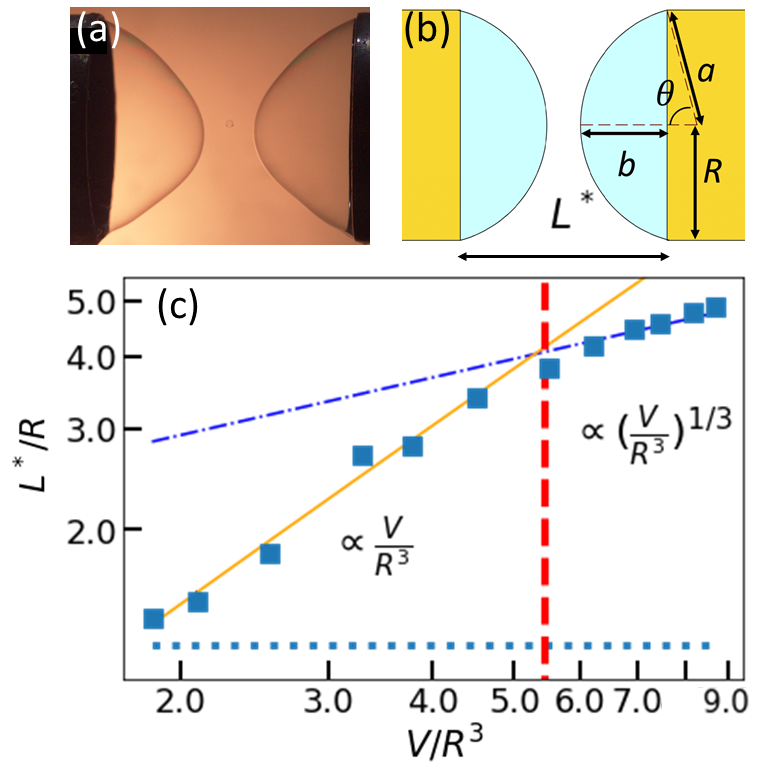}
	
    \caption{ (a) Bubble was split in half after pinch-off. (b) Illustration for how $L^*=5b/2$ is determined by the boundary conditions. (c) Blue squares show the experimental results in full-log plot for dimensionless critical length ${L^{*}}/{R}$ vs. volume ${V}/{R^{3}}$. The red dashed line ${V}/{R^{3}}=5.44$ separates two regions with different exponents, 1 and 1/3 for the orange and purple dash-dotted lines and blue dotted line indicates the value of $L^*$ for film.}
	
    \label{fig:lvp3}

\end{figure}

\subsection{Breakup regime}

Different from the   flat surfaces on rings for film, two spherical bubbles survive the breakage and appear on the caps for bubble. The spherical shape is to minimize the surface energy, for which the height $b$ defined in Fig. \ref{fig:lvp3} (a, b)  can be calculated. In the mean time,  the critical length $L^{*}$ at which the irreversible processes are initiated can also be determined theoretically by the breakdown of solution from minimizing the potential energy for  equilibrium regime. Comparing these two lengths, we find that $L^{*}$ is not only bigger than $2b$, which explains the necessity for both bubbles to retract and breakage, but roughly equals $5b/2$. This is verified in our experiment.

Depending on the amount of $V$, we expect two scenarios for the remnant bubble. Straightforward calculation in Sec. IV C reveals that   $L^{*}/R\propto (V/R^{3})^\beta$  with   $\beta =1$ for $V/R^{3} \ll 5.44$ and  $1/3$ otherwise. This prediction is nicely verified by   Fig.~\ref{fig:lvp3}(c).

\section{Theoretical derivation}
\subsection{Equilibrium regime}

Experimentally we can stretch bubble and film horizontally or vertically. Although the sagging and non-symmetric contour due to  gravity  in both cases  can be alleviated by minimizing $V$, it is still uncertain whether the critical behavior at pinch-off will be affected. So theoretical calculations can help us not only clarify this concern, but also give us analytic expressions for quantities of our interest, e.g., $h_{\rm min}$ and highlight the influence of volume conservation.  To realize how the parameters affect the contour of bubble, we start from the minimization of total energy for bubble:
\begin{eqnarray}
\frac{U}{2} = \int_{0}^{L/2}\left[\gamma \cdot 2\pi h\sqrt{1+h'^{2}} + \lambda \left(\pi h^{2}-\frac{V}{L} \right)\right]dx
\label{energy}
\end{eqnarray}
where  the Lagrange multiplier $\lambda$ makes sure that the air volume is conserved.  Using the second form of Euler-Lagrange equation, we can obtain
\begin{eqnarray}
\frac{h}{\sqrt{1+h'^{2}}}+\frac{\lambda}{2\gamma}h^{2} = h_{\rm {min}}+ \frac{\lambda}{2\gamma}h_{\rm{min}}^{2}. 
\label{eq:euler}
\end{eqnarray}
After some transpositions, Eq. (\ref{eq:euler}) becomes
\begin{eqnarray}
h'=\sqrt{\left(\frac{h}{h_{\rm{min}}+\frac{\lambda}{2\gamma}h_{\rm{min}}^{2}-\frac{\lambda}{\gamma}h^{2}} \right)^{2}-1}.
\end{eqnarray}
Solving this differential equation will enable us to obtain information of the contour $h(x)$:
\begin{eqnarray}
x=\bigints_{h_{\rm{min}}}^{h} dh/\sqrt{\left(\frac{h}{h_{\rm{min}}+\frac{\lambda}{2\gamma}h_{\rm{min}}^{2}-\frac{\lambda}{\gamma}h^{2}} \right)^{2}-1}.
\end{eqnarray}
where $0\le x \le L/2$. 

By implementing the boundary condition that $h(x=L/2)=R$ and volume conservation, we get
\begin{eqnarray}
\frac{L}{2h_{\rm{min}}} = \bigints_{1}^{\frac{R}{h_{\rm{min}}}}{dy}/{\sqrt{\Big(\frac{y}{1+ \big(1-y^{2}\big)\xi}\Big)^{2}-1}}
\label{length_f}
\end{eqnarray} 
and
\begin{eqnarray}
\frac{V}{2h_{min}^{3}} = \bigints_{1}^{\frac{R}{h_{\rm{min}}}}{\pi y^{2}dy}/{\sqrt{\Big(\frac{y}{1+ \big(1-y^{2}\big)\xi}\Big)^{2}-1}}.
\label{volume_f}
\end{eqnarray}
where a change of variable $y=h/h_{\rm{min}}$ has been performed to render the parameters dimensionless and $\xi \equiv \frac{\lambda}{2\gamma}h_{\rm{min}}$.
By setting $\lambda =0$, Eq. (\ref{length_f}) will revert to depicting a film and give us $h_{\rm{min}}/R$ vs. $L/R$ in agreement with Fig.~\ref{fig:sp_volume}(a). During the stretching of bubble, there must be a period when $h_{\rm{min}}$ is close to $R$ and we can approximate $h_{\rm{min}}$ by $R- \Delta h$ where $\Delta h \ll R$. This allows us to Taylor expand $R/h_{\rm{min}} = R/(R-\Delta h) \approx 1+ \Delta h/R$, Eqs. (\ref{length_f}) and (\ref{volume_f}) to get
\begin{eqnarray}
\frac{L}{2 h_{\rm{min}}} &\approx& \frac{\left(\frac{R}{h_{\rm{min}}}-1\right)}{\sqrt{\big(\frac{1+ \frac{\Delta h}{R}}{1+\big(1-\frac{\Delta h}{R}\big)^{2}\xi}\big)^{2}-1}} \nonumber \\
&=& \frac{\Delta h/R}{\sqrt{\big(\frac{1+\Delta h/R}{1-2\xi {\Delta h}/{R}}\big)^{2}-1}}
\label{length_taylor}
\end{eqnarray}
and
\begin{eqnarray}
\frac{V}{2h_{\rm{min}}^{3}} \approx \frac{\Delta h/R}{\sqrt{\big(\frac{1+\Delta h/R}{1-2\xi {\Delta h}/{R}}\big)^{2}-1}}\big(1+2\Delta h/R\big)
\label{volume_taylor}
\end{eqnarray}
where terms of order higher than ${\Delta h}/{R}$ have been neglected. 
By comparing the above two equations, we obtain $V/(2\pi h_{\rm min}^{3})\approx {L}/{2 h_{\rm{min}}}$. Note that this result predicts a simple yet informative relation for how $h_{\rm min}$ varies with $L$:
\begin{eqnarray}
h_{\rm min} \approx \sqrt{\frac{V}{\pi L}}.
\label{cylinder}
\end{eqnarray}
which matches the data in early equilibrium regime for Fig.~\ref{fig:sp_volume}(b). In retrospect, this result is expected from our restricting $h_{\rm{min}}\approx R$  since the shape of bubble now mimics that of a cylinder. 

The derivations from Eq. (\ref{eq:euler}) to (\ref{volume_taylor}) serve several purposes: First, to illustrate the important role of the Lagrange multiplier $\lambda$ and how it mathematically prohibits the bubble from adopting the catenoid profile as the film. Second, $h''$ near the edge can be easily shown to exhibit different signs for film  and bubble - positive and negative from the derivative of Eq. (\ref{eq:euler}), respectively, as detailed in Sec. X of the SM \cite{sm}. Third, they highlight the importance of $\lambda$ and subsequent parameter $\xi\approx [-2+(\Delta h/R)]/4$, without which  the right-hand side of Eq. (\ref{length_taylor}) will be of order $\mathcal{O}(\sqrt{\Delta h/R})$ - meaning the cylindrical shape is only possible when the two rings  are very close for film. 

\begin{figure}[h!]
	\centering
	\includegraphics[width=6cm]{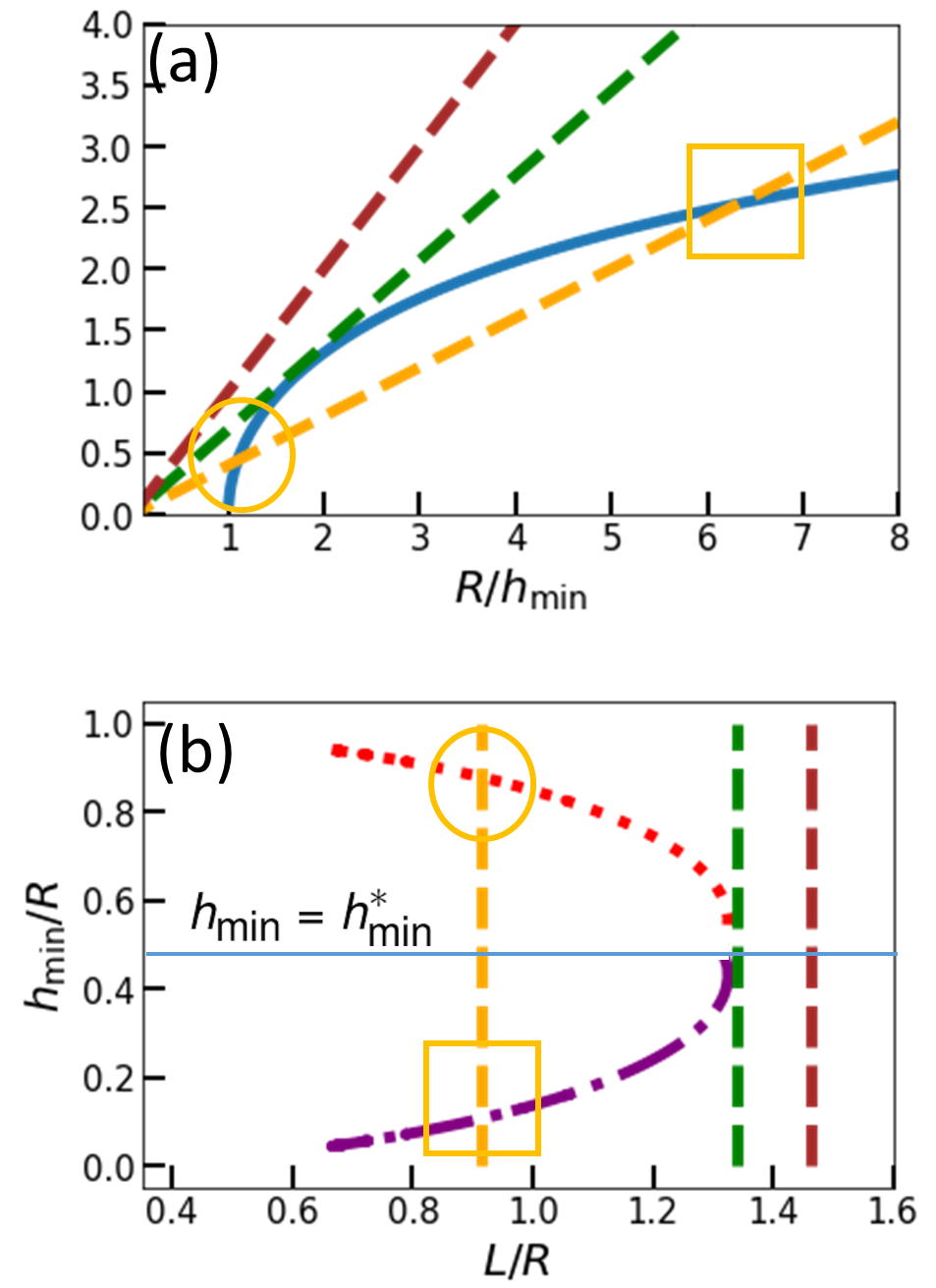}
	
    \caption{(a) The right- and left-hand sides of  Eq. (\ref{length_f}) with $\xi =0$ are plotted in solid and dashed lines as a function of $R/h_{\rm{min}}$.   Three scenarios are possible by increasing  the slope $L/(2R)$ with zero, one, and two interceptions that are denoted respectively by the brown, green and yellow dashed lines.  The green line defines the critical $L^*$ and $h^*_{\rm{min}}$. In the mean time, since we expect $h_{\rm min}$ to shrink as $L$ lengthens, the solution highlighted by the yellow square should be discarded.  This unphysical solution is further represented by the purple dash-dot line in (b) that shows $h_{\rm{min}}/R$ vs. $L/R$. }
    \label{fig:eq_op_reverse}

\end{figure}

Some experience can be borrowed from the analysis of film. After setting $\xi=0$,  we plot both sides of  Eq. (\ref{length_f})  as a function of  $R/h_{\rm{min}}$ in Fig. \ref{fig:eq_op_reverse} (a).
When $L>L^*$, there is no intersection - meaning that the starting point of minimizing the surface energy is problematic. This is consistent with our expectation that film will collapse automatically at large $L$ when we need to resort to minimizing the action. When $L=L^*$,  both sides of the equation become tangent. The same is expected for bubble, i.e., we should differentiate both sides of Eq. (\ref{length_f}) with respect to $h_{\rm{min}}$  to locate $L^*$. There are two intersections for $L<L^*$ with one being unphysical as explained in Fig. \ref{fig:eq_op_reverse} (a).  When we plot $h_{\rm{min}}$ vs. $L$, the solution should be double-valued until $L$ reaches $L^*$  in Fig. \ref{fig:eq_op_reverse}(b). Therefore, we expect $L(h_{\rm{min}})\approx -\chi (h_{\rm{min}}-h^*_{\rm{min}})^2 +L^*$ where $\chi$ is a constant when $h_{\rm{min}}$ is close to  the critical neck radius $h^*_{\rm{min}}$ beyond which the neck collapses spontaneously. Simple rearrangement gives 
\begin{eqnarray}
h_{\rm{min}} \approx h^*_{\rm{min}}+ \chi^{-1/2} \sqrt{L^{*}-L}.
\label{bubble_equillbrium_end}
\end{eqnarray}
Our confidence on Eqs. (\ref{cylinder}) and (\ref{bubble_equillbrium_end}) is supported by its rightful prediction of  an inflection point in Fig. \ref{fig:sp_volume}(b) due to the  fact that their curvatures are of opposite sign.

\subsection{Breakup regime}
The collapsing speed of neck is very fast. As will be delineated later, there  will be two complications before the final breakage. First, two necks will be developed in pinch-off regime. Second, this is followed by the breaking stage when a satellite bubble is formed in the middle of these two necks after the hollow thin tube connecting them becomes a liquid string. There is no gas leakage throughout breaking and relaxing stages and the volume of satellite bubble can be neglected. Therefore, $V$ should equal to the combined volume of the two remnant bubbles after breakage. We can directly obtain the relation between $V$ and $L^{*}$ which can be estimated from Fig. \ref{fig:lvp3}(b) where the chord length equals $2R$ and the radius of partial sphere is denoted by $a$. From the geometry in Fig. \ref{fig:lvp3}(b), we can write down
\begin{equation}
a=\frac{R^{2}+b^2}{2b}
\label{radius}
\end{equation}
and 
\begin{equation}
\cos\theta = \frac{R^{2}-b^2}{R^{2}+b^2}.
\label{angle}
\end{equation}
The volume of each partial sphere can be calculated as
\begin{eqnarray}
\frac{V}{2}= \frac{\pi L^{*}}{15}\left(3R^{2}+b^2\right)
\label{volume_dome}
\end{eqnarray}
Rearranging both sides to make them dimensionless, we found 
\begin{eqnarray}
\frac{V}{R^{3}}=\frac{\pi}{3}\frac{b}{R}\Big[3+\Big(\frac{b}{R}\Big)^{2}\Big]
\label{vd_dimensionless}
\end{eqnarray}
By inserting the experimental result $b=2L^*/5$, there are two limiting cases to Eq. (\ref{vd_dimensionless}).
When ${L^{*}}/{R} \ll \frac{5}{2} \sqrt{3}$, the cubic term in Eq. (\ref{vd_dimensionless}) can be neglected and
\begin{eqnarray}
\frac{V}{R^{3}} \cong \frac{\pi}{2}\frac{L^{*}}{R}
\label{eq:small V}
\end{eqnarray}
In the other extreme ${L^{*}}/{R} \gg \frac{5}{2}  \sqrt{3}$, the linear term becomes negligible and
\begin{eqnarray}
\frac{V}{R^{3}} \cong \frac{8\pi}{375} \left(\frac{L^{*}}{R} \right)^{3}.
\label{eq:big V}
\end{eqnarray}
The two regions in Eqs. (\ref{eq:small V}) and (\ref{eq:big V}) are vindicated by Fig. \ref{fig:lvp3}(c).

\section{conclusion and discussions}
 We studied how the mediation of long-range pressure that comes in via the volume $V$ conservation affects the necking phenomenon  of soap bubble, as compared to  the relatively well-studied film. Understandably the  distinct contour shape exhibited by bubble should result in different ways for how the  neck radius $h_{\rm min}$ shrinks with increasing separation length in equilibrium regime.
Upon entering   non-equilibrium state, film and  bubble share the following   properties: (1) the collapsing dynamics is insensitive to the pulling speed $v_{s}$ and ring or cap size $R$, (2) the contour for roll-off  regime (a)  exhibits self-similarity in its evolution  which can be quantified by  cosine similarity and supported by theoretic derivations and (b)  is universal, i.e., independent of $R$ and $V$ upon being rescaled by $h_{\rm min}$ - although varying $V/R^{3}$ will render a different universal line for bubble which elaborates the effect of erasing the boundary condition is earlier than   pinch-off regime.   
 (3) The evolution of $h_{\rm{min}}\propto \tau^{2/3}$ is independent of $v_{s}$, $R$ and $V$ for pinch-off regime. 
However, the  $h_{\rm{min}} \propto \tau^{\alpha}$ relation is found to differ for  roll-off regime with $\alpha =1/2$ for bubble, in contrast to 2/3 for film. 

The dimensionless threshold length $L^{*}/R$ that marks the beginning of spontaneous collapse for bubble is found to depend only on the ratio of $V$ and $R^3$ and can be separated into two different regimes for which  a simple theory was built.

We suggest that future researchers can strengthen the following aspect:
To determine whether  $\alpha$ will be affected by $v_{s}$  when it is comparable to the collapse speed  $\bar{v_{c}}$ in roll-off regime.  The motivation is that most  breakage studies concentrate on pinch-off regime when   $\bar{v_{c}}$ is so fast that it is challenging to apply a  fast  $v_{s}$  without invoking  unwanted   technical artifacts. One exception, though, is Ref.\cite{universal_stone} where $\bar{v_{c}}$ was slowed down considerably by the viscous stress  and $h_{\rm min}$ was found to scale as $\tau^{1/5}$ in early self-similar regime. It gave the authors an incentive to probe how a large enough  $v_{s} \sim \bar{v_{c}}$ affects this power law. Surprisingly, the result turns out to be negative.  Likewise, it is recommended to increase  $v_s$ to the same order as $\bar{v_{c}}$ in our roll-off regime.  

We are grateful to  C. Y. Lai and J. R. Huang for  useful discussions and thank P. Yang and J. C. Tsai for the use of high-speed cameras. Financial support from the Ministry of Science and Technology in Taiwan under Grants No. 105-2112-M007-008-MY3 and No. 108-2112-M007-011-MY3 is acknowledged.

\appendix

\section{Evidence for $V$-independent $h_{\rm min}(\tau )$ in roll-off regime}
\begin{figure}[h!]
	\includegraphics[width=7.cm]{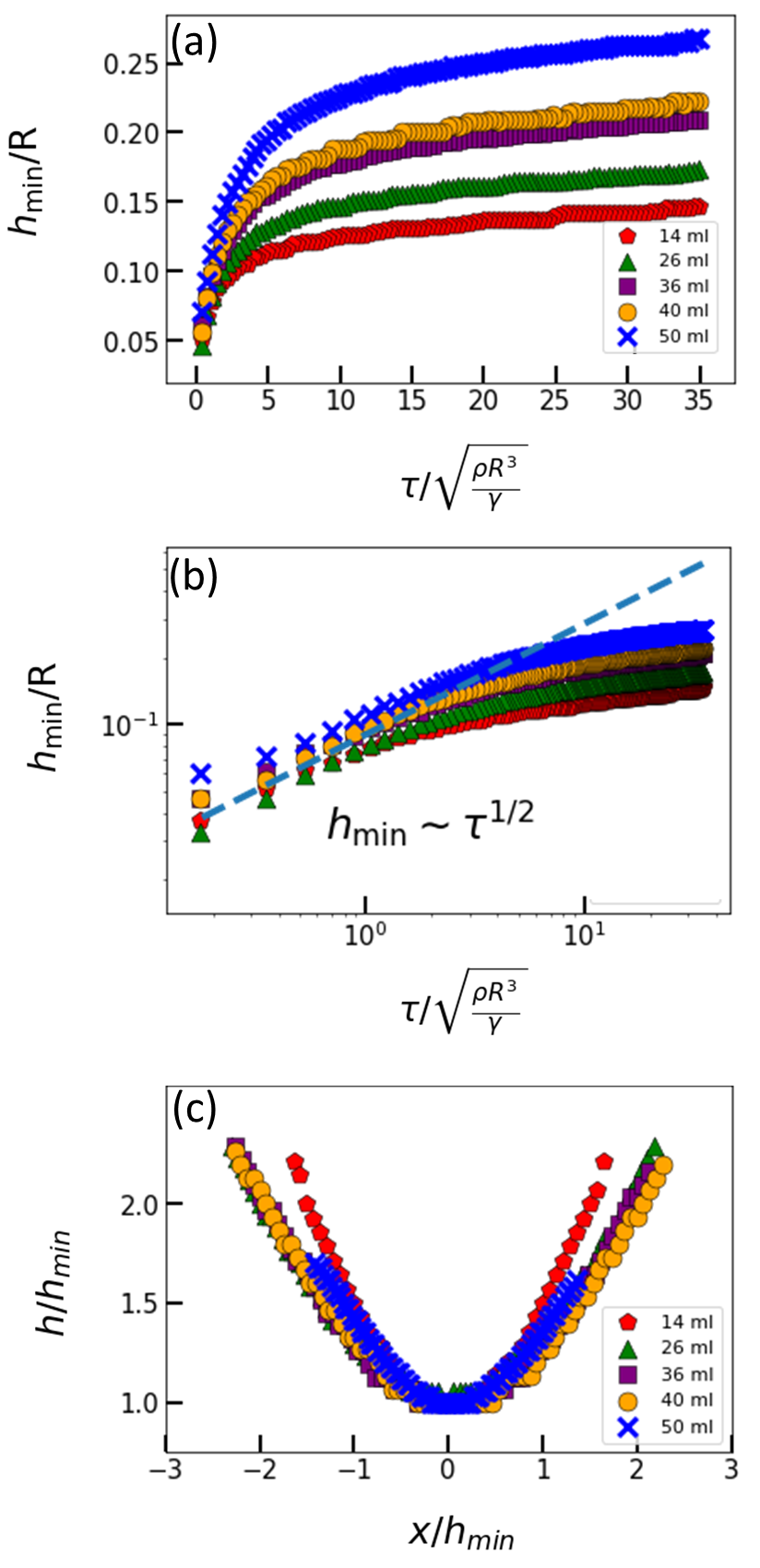}
	
    \caption{ Rescaled $h_{\rm{min}}/R$ is plotted against  $\tau /\sqrt{\frac{\rho R^3}{\gamma}}$  for bubbles of different $V$ in (a) linear and (b) log-log plot where $R$ = 20 mm and $v_{s}= 16$ mm/s. (c) With the exception of small $V$= 14 ml, overlap of master curves for different $V$ indicates that they are almost identical. }

    \label{fig:volume_appendix_tot}		
    
\end{figure}

It is obvious that $h_{\rm{min}}$ will increase with  $V$ when all other parameters are fixed in  equilibrium regime. Interestingly, it turns out that the properties characteristic of   roll-off regime become insensitive to the actual value of $V$ in Fig. \ref{fig:volume_appendix_tot} (a, b), indicating that this is a local event in which very little air is involved.  Proof can be found in  the rescaled profile  in Fig. \ref{fig:volume_appendix_tot}(c) where an extra influx of $V$ is shown to pile up  on the flank of, but not at the  neck. 

\bibliographystyle{plainnat}

\bibliography{apssamp}

\end{document}